\documentclass[letterpaper,twocolumn,10pt]{article}
\usepackage{usenix2019_v3,epsfig}
\usepackage{epigraph}
\usepackage{makecell} 
\usepackage{multirow}
\usepackage{xspace}
\usepackage{paralist}
\usepackage{libertine}
\usepackage{eurosym}
\usepackage{amssymb}
\usepackage{xcolor}
\usepackage{graphicx}

\newcommand{\sref}[1]{\S\ref{#1}}
\newcommand{\vheading}[1]{\vspace{0.05in}\noindent\textbf{#1}}

\begin{document}

\date{}

\title{\LARGE \bf The Seven Sins of\\ Personal-Data Processing Systems Under GDPR}

\author{
{\normalfont Supreeth Shastri}\\
{\normalsize Computer Science}\\
{\normalsize University of Texas at Austin}
\and
{\normalfont Melissa Wasserman}\\
{\normalsize School of Law}\\
{\normalsize University of Texas at Austin}
\and
{\normalfont Vijay Chidambaram}\\
{\normalsize Computer Science}\\
{\normalsize University of Texas at Austin}
} 

\maketitle

\pagestyle{empty}

\subsection*{Abstract}

In recent years, our society is being plagued by unprecedented levels of privacy and security breaches. To rein in this trend, the European Union, in 2018, introduced a comprehensive legislation called the General Data Protection Regulation (GDPR). In this paper, we review GDPR from a system design perspective, and identify how its regulations conflict with the design, architecture, and operation of modern systems. We illustrate these conflicts via \emph{the seven GDPR sins}: storing data forever; reusing data indiscriminately; walled gardens and black markets; risk-agnostic data processing; hiding data breaches; making unexplainable decisions; treating security as a secondary goal. Our findings reveal a deep-rooted tussle between GDPR requirements and how modern systems have evolved. We believe that achieving compliance requires comprehensive, grounds up solutions, and anything short would amount to \emph{fixing a leaky faucet in a sinking ship}.

\section{Introduction}

Modern computing systems exhibit unprecedented levels of scalability, 
reliability, and affordability. For example, Amazon's cloud 
computing infrastructure provides on-demand access to inexpensive 
computing to over 1 million users in 190 countries, all the while 
guaranteeing four nines of availability. Similarly, Google operates 
8 global-scale applications at 99.99\% uptime with each of them
supporting more than 1 billion users. As Internet-era systems 
focus on performance, cost-efficiency, reliability, 
and scalability as their primary design goals, security and privacy 
have taken a backseat. 

However, it was not until recently that we realized the impact 
of relegating data security and user privacy as afterthoughts in 
system design. In 2013, Yahoo! experienced a theft of 3 billion user
records; in 2016, Facebook had its user data illegally harvested to 
influence the U.S. and U.K. democratic processes; equally worse, 
it was discovered that many companies were indiscriminately collecting 
and using personal data without people's consent. In response to 
these developments, the European Union (EU) adopted 
a privacy regulation called the General Data Protection 
Regulation (GDPR) \cite{gdpr-regulation}. By defining the privacy 
of personal data as a fundamental right of all European people, 
GDPR regulates the lifecycle of personal data. Thus, any 
company dealing with EU customers is legally bound to comply with 
GDPR. 

In this work, we examine how GDPR affects the design and operation 
of modern computing systems. Surprisingly, our analysis reveals that
several design principles and architectural elements of real-world 
systems are at odds with the proposed regulation. We highlight seven 
such principles and practices, \emph{the seven GDPR sins}, by
discussing how they came to be, reviewing the conflicting regulation, and
chronicling their real-world implications. For example, given the 
commercial value of personal data, modern systems naturally evolved 
to store them forever, to reuse them across various applications, 
to sell them for profit, and to stash them in walled gardens. However,
GDPR either explicitly forbids or severely restricts the scope of all 
these practices.
  
The goal of this paper is three folds: first, we provide a brief 
primer on GDPR (\sref{sec-gdpr}). Next, we illustrate the tussle 
between GDPR requirements and modern systems (\sref{sec-design}). 
Finally, we shed light on the challenges of retrofitting existing 
systems into compliance (\sref{sec-discussion}).

\section{GDPR}
\label{sec-gdpr}

On May 25th 2018, the European parliament adopted the General Data Protection Regulation~\cite{gdpr-regulation}. In contrast with targeted privacy regulations like HIPAA~\cite{hipaa} and FERPA~\cite{ferpa}, GDPR takes a comprehensive view by defining \emph{personal data} to be any information relating to an identifiable natural person. Then, GDPR defines three entities that interact with personal data: (i) \emph{data subject}, the person whose personal data is collected, (ii) \emph{data controller}, the entity that collects and uses personal data, and (iii) \emph{data processor}, the entity that processes personal data on behalf of a data controller. Consider the music streaming company Spotify collecting its customer's listening history, and then using Google cloud's services to identify new recommendations for customers. In this scenario, Spotify is the data controller and Google Cloud is the data processor. In the following, we provide a brief background on GDPR regulations and their impact. 

\vheading{Structure}.
GDPR is organized as 99 \emph{articles} that describe its legal requirements, and 173 \emph{recitals} that provide additional context and clarifications to these articles. The first 11 articles layout the principles of data privacy; articles 12-23 establish the rights of the people; then articles 24-50 mandate the responsibilities of the data controllers and processors; the next 26 articles describe the role and tasks of supervisory authorities; and the remainder of the articles cover liabilities, penalties and specific situations. We expand on the relevant articles in \sref{sec-design}.

\vheading{Impact}. 
Compliance with GDPR has been a challenge for technology companies. A number of companies like Instapaper, Klout, and Unroll.me completely terminated their services in Europe to avoid the hassles of compliance. Few other businesses made temporary modifications. For example, media site \emph{USA Today} turned off all advertisements~\cite{usa-today-no-ads}, whereas \emph{the New York Times} stopped serving personalized ads~\cite{nytimes-no-ads}. While most organizations are working towards compliance, Gartner reports~\cite{gartner-prediction} that less than 50\% of the companies affected by GDPR were compliant by the end of 2018. This challenge is further exacerbated by the performance impact that GDPR-compliance imposes on current systems~\cite{gdpr-storage}.

In contrast, people have been enthusiastically exercising their newfound rights, and not been shy to report any shortcomings. In fact, the EU data protection board reports~\cite{gdpr-in-numbers} having received 95,180 complaints from individuals and organizations in the first 8 months of GDPR. Surprisingly, even the companies have been forthcoming in reporting their security failures and data breaches, with 41,502 breach notifications sent to regulators in the same 8 month period. 

\vheading{In the cloud}. GDPR brings two distinct challenges to cloud computing. First, as cloud has become the de-facto computing platform for the modern society, companies and organizations have to rely on cloud services to realize compliance at their application level. In fact, as we discuss in \sref{sec-sin7}, GDPR precludes companies from using those cloud providers who do not support compliance efforts. Second, the large-scale Internet-era systems that constitute (and drive) the cloud are themselves at odds with several GDPR regulations. Thus, cloud providers must face the compliance challenges, both externally and internally.

\section{The Seven GDPR Sins}
\label{sec-design}

Many of the design principles, architectural components, and operational practices of modern computing systems conflict with the rights and responsibilities laid out in GDPR. We discuss seven such practices below.

\subsection{Storing Data Forever} 
Computing systems have always relied on insights derived from data. However, this dependence is reaching new heights, especially in this decade, with widespread adoption of machine learning and big data analytics in system design. Data has been compared to oil, electricity, gold, and even bacon~\cite{data-is-bacon}. Naturally, technology companies evolved to not only collect user data aggressively but also to preserve them forever. However, GDPR mandates that no data lives forever.

\begin{quote} 
\textsc{Article 17: Right to be forgotten.} \textsl{``(1) The data subject shall have the right to obtain from the controller the erasure of personal data without undue delay [...]''}
\end{quote}

\begin{quote}
\textsl{Article 13: \textsc{Information to be provided where personal data are collected from the data subject.} ``(2)(a) [...] the controller shall provide the period for which the personal data will be stored, or the criteria used to determine that period;''}
\end{quote}

\begin{quote} 
\textsl {Article 5(1)(e): \textsc{Storage limitation.} ``[...] kept for no longer than is necessary for the purposes for which the personal data are processed [...]''}
\end{quote}

GDPR grants users an unconditional right, via article 17, to request their personal data be removed from everywhere in the system within a reasonable time. In conjunction with this, articles 5 and 13 lay out additional responsibilities for the data controller: (i) at the point of collection, users should be informed the time period for which their personal data would be stored, and (ii) if the personal data is no longer necessary for the purpose for which it was collected, then it should be deleted. These simply mean that all personal data should have a time-to-live (TTL) that users are aware of, and that controllers honor. However, this restriction does not apply to archiving in the public interest, or for scientific or historical research purposes. 

\vheading{Deletion in the real-world}. While conceptually clear, a timely and guaranteed removal of data is challenging in practice. For example, Google cloud describes the deletion of customer data as an iterative process~\cite{google-deletion} that could take up to 180 days to fully complete. This is because, for performance, reliability, and scalability reasons, parts of data gets replicated in various storage subsystems like memory, cache, disks, tapes, and network storage; multiple copies of data is saved in redundant backups and geographically distributed datacenters. Such practices not only delay the timeliness of deletions but also make it harder to guarantee it. 

\subsection{Reusing Data Indiscriminately}

While designing software systems, a purpose is typically associated with programs and models, whereas data is viewed as a helper resource that serves these high-level entities in accomplishing their goals. This portrayal of data as an inert entity allows it to be used freely and fungibly across various systems. For example, this has enabled organizations like Google and Amazon to collect user data once, and use it to personalize their experiences across several services. However, GDPR regulations prohibit this practice. 
 
\begin{quote} 
\textsc {Article 5(1)(b): Purpose limitation.} \textsl{``Personal data shall be collected for specified, explicit and legitimate purposes and not further processed in a manner that is incompatible with those purposes [...]''}
\end{quote}

\begin{quote} 
\textsc {Article 6: Lawfulness of processing.} \textsl{``(1)(a) Processing shall be lawful only if [...] the data subject has given consent to the processing of his or her personal data for one or more specific purposes.''}
\end{quote}

\begin{quote} 
\textsc {Article 21: Right to object.} \textsl{``(1) The data subject shall have the right to object at any time to processing of personal data concerning him or her [...]''}
\end{quote}

The first two articles establish that personal data could only be collected for specific purposes and not be used for anything else. Then, article 21 grants users a right to object, at any time, to their personal data being used for any purpose including marketing, scientific research, historical archiving, or profiling. Together, these articles require each personal data item to have its own blacklisted and whitelisted purposes that could be changed over time. 

\vheading{Purpose in the real-world}. The impact of the purpose requirement has been swift and consequential. For example, in January 2019, the French data protection commission~\cite{google-purpose-bundling} fined Google \texteuro50M for not having a legal basis for their ads personalization. Specifically, the ruling said that the user consent obtained by Google was not ``specific'' enough, and the personal data thus obtained should not have been used across 20 services.

\subsection{Walled Gardens and Black Markets}
As we are in the early days of large-scale commoditization of personal data, the norms for acquiring, sharing, and reselling them are not yet well established. This has led to uncertainties for people and a tussle for control over data amongst controllers. People are concerned about vendor lock-ins, and about a lack of visibility once their data is resold or shared in the secondary markets. Organizations have responded to this by setting up walled gardens, and making secondary markets more opaque. However, GDPR dismantles such practices.

\begin{quote} 
\textsc {Article 20: Right to data portability.} \textsl{``(1) The data subject shall have the right to receive the personal data concerning him or her, which he or she has provided to a controller. (2) [...] the right to have the personal data transmitted directly from one controller to another.''}
\end{quote}

\begin{quote} 
\textsc {Article 14: Information to be provided where personal data have not been obtained from the data subject.} \textsl{``(1) (c) the purposes of the processing [...], (e) the recipients [...], (2) (a) the period for which the personal data will be stored [...], (f) from which source the personal data originate [...]. (3) The controller shall provide the information at the latest within one month.''}
\end{quote}

With article 20, people have a right to request for all the personal data that a controller has collected directly from them. Not only that, they could also ask the controller to directly transmit all such personal data to a different controller. While that tackles the vendor lock-ins, article 14 regulates the behavior in secondary markets. It requires that anyone indirectly procuring personal data must inform the users, within a month, about (i) how they acquired it, (ii) how long would they be stored, (iii) what purpose would they be used for, and (iv) who they intend to share it with. The \emph{data trail} set up by this regulation should bring the control and clarity back to the people.

\vheading{Data movement in the real-world.} When GDPR went live, a large number of companies rolled out~\cite{gdpr-data-download} data download tools for EU users. For example, Google Takeout~\cite{google-takeout} lets users not only access all their personal data in their system but also port data directly to external services. However, the impact has been less savory for programmatic ad exchanges~\cite{gdpr-programmatic-ad-buy} in Europe, many of which had to shut down. This was primarily due to Google and Facebook restricting access to their platforms for those ad exchanges, which could not verify the legality of the personal data they possessed.

\subsection{Risk-Agnostic Data Processing}

Modern technology companies face the challenge of creating and managing increasingly complex software systems in an environment that demands rapid innovation. This has led to a practice, especially in the Internet-era companies, of prioritizing speed over correctness; and to a belief~\cite{move-fast-break-things} that \textit{unless you are breaking stuff, you are not moving fast enough}~\cite{zuckerberg-quote}. However, GDPR explicitly restricts this approach when dealing with personal data. 

\begin{quote} 
\textsc {Article 35: Data protection impact assessment.} \textsl{``(1) Where processing, in particular using new technologies, is likely to result in a high risk to the rights and freedoms of natural persons, the controller shall, prior to the processing, carry out an assessment of the impact of the envisaged processing operations on the protection of personal data.''}
\end{quote}

\begin{quote} 
\textsc {Article 36: Prior consultation.} \textsl{``(1) The controller shall consult the supervisory authority prior to processing where [...] that would result in a high risk in the absence of measures taken by the controller to mitigate the risk.''}
\end{quote}

GDPR establishes, via articles 35 and 36, two levels of checks for introducing new technologies and for modifying existing systems, if they process large amounts of personal data. The first level is internal to the controller, where an impact assessment must analyze the nature and scope of the risks, and then propose the safeguards needed to mitigate them. Next, if the risks are systemic in nature or concern common platforms, either internal and external, the data protection officer must consult with the supervisory authority prior to any processing.

\vheading{Fast and broken in the real-world}. Facebook, despite having moved away from the aforementioned motto, has continued to be plagued by it. In 2018, it revealed two major breaches: first, that their APIs allowed Cambridge Analytica to illicitly harvest~\cite{cambridge-analytica} personal data from 87M users, and then their new \emph{View As} feature was exploited~\cite{facebook-view-as-leak} to gain control over 50M user accounts. However, this practice of prioritizing speed over security is not limited to one organization. For example, in Nov 2017, fitness app Strava released an athlete motivation tool called global heatmap~\cite{strava-heatmap} that visualized athletic activities of worldwide users. However, within months, these maps were used to identify undisclosed military bases and covert security operations~\cite{strava-military-leak}, jeopardizing missions and lives of soldiers. 

\subsection{Hiding Data Breaches}

The notion that one is \emph{innocent until proven guilty} predates all computer systems. As a legal principle, it dates back to 6th century Roman empire~\cite{roman-law}, where it was codified that \emph{proof lies on him who asserts, not on him who denies}. Thus, in the event of a data breach or a privacy violation, organizations typically claim innocence and ignorance, and seek to be absolved of their responsibilities. However, GDPR makes such presumption conditional on the controller proactively implementing risk-appropriate security measures (i.e., accountability), and notifying breaches in a timely fashion (i.e., transparency).

\begin{quote} 
\textsc {Article 5: Principles relating to processing.} \textsl{``(1) Personal data shall be processed with [...] lawfulness, fairness and transparency; [...] purpose limitation; [...] data minimisation; [...] accuracy; [...] storage limitation; [...] integrity and confidentiality. (2) The controller shall be responsible for, and be able to demonstrate compliance with (1).''}
\end{quote}

\begin{quote} 
\textsc {Article 33: Notification of a personal data breach.} \textsl{``(1) the controller shall without undue delay and not later than 72 hours after having become aware of it, notify the supervisory authority. [...] (3) The notification shall at least describe the nature of the personal breach, [...] likely consequences, and [...] measures taken to mitigate its adverse effects.''}
\end{quote}

GDPR's goal is two folds: first, to reduce the frequency and impact of data breaches, article 5 lays out several ground rules. The controllers are not only expected to adhere to these internally but also be able to demonstrate their compliance externally. Second, to bring transparency in handling data breaches, articles 33 and 34 mandate a 72 hour notification window within which the controller should inform both the supervisory authority and the affected people. 

\vheading{Data breaches in the real-world.} In recent years, responses to personal data breaches have been adhoc: while a few organizations have been forthcoming, others have chosen to refute~\cite{aadhar-refute-breach}, delay~\cite{panera-delay-breach} or even pay off hackers~\cite{uber-breach-payoff}. However, GDPR's impact has been swift and clear. Just in the first 8 months (May 2018 to Jan 2019), regulators have received 41,502 data breach notifications~\cite{gdpr-in-numbers}. This number is in stark contrast from the pre-GDPR era, with reports~\cite{pre-gdpr-numbers} of 945 worldwide data breaches in the first half of 2018. 

\subsection{Making Unexplainable Decisions}
Algorithmic decision-making has been successfully applied to several domains: curating media content, managing industrial operations, trading financial instruments, personalizing advertisements, and even combating fake news. Their inherent efficiency and scalability (with no human in the loop) has made them a necessity in modern system design. However, GDPR takes a cautious view of this trend.

\begin{quote} 
\textsc {Article 22: Automated individual decision-making.} \textsl{``(1) The data subject shall have the right not to be subject to a decision based solely on automated processing [...]''}
\end{quote}

\begin{quote} 
\textsc {Article 15: Right of access by the data subject.} \textsl{``(1) The data subject shall have the right to obtain from the controller [...] meaningful information about the logic involved, as well as the significance and the envisaged consequences of such processing.''}
\end{quote}

On one hand, privacy researchers from Oxford postulate~\cite{gdpr-explanation} that these two regulations, together with recital 71, establish a ``right to explanation'' and thus, human interpretability should be a design consideration for machine learning and artificial intelligence systems. However, another group at Oxford argues~\cite{gdpr-no-right-to-explanation} that GDPR falls short of mandating this right by requiring users (i) to demonstrate significant consequences, (ii) to seek explanation only after a decision has been made, and (iii) to have to opt out explicitly.

\vheading{Decision-making in the real-world.} The debate over the privacy and interpretability in automated decision-making has just begun. Starting 2016, the machine learning and intelligence community began exploring this rigorously: the workshop on Explainable AI~\cite{xAI} at IJCAI, and the workshop on Human Interpretability in Machine Learning~\cite{WHI} at ICML being two such efforts. In January 2019, privacy advocacy group NoYB has filed~\cite{noyb-streaming-service-complaints} complaints against eight streaming services including Amazon, Apple Music, Netflix, SoundCloud, Spotify, YouTube, Flimmit and DAZN for violating the article 15 requirements in their recommendation systems. 

\subsection{Security as Secondary Goal}
\label{sec-sin7}
The functionality-first approach is not specific to modern computing systems, rather it permeates through much of the computing history. For example, the Internet, which forms the foundation for cloud computing, was never designed with security in mind. It also illustrates the difficulties of retrofitting a functional system with afterthought security. Combating this practice is one of the central tenets of GDPR. 

\begin{quote} 
\textsc {Article 25: Data protection by design and by default.} \textsl{``(1) [...] design to implement data-protection principles in an effective manner. (2) [...] ensure that by default, only personal data which are necessary for each specific purpose are processed, and [...] personal data are not made accessible to an indefinite number of persons.''}
\end{quote}

\begin{quote} 
\textsc {Article 24: Responsibility of the controller.} \textsl{``the controller shall implement appropriate technical and organisational measures to ensure, and to be able to demonstrate that processing is performed in accordance with this Regulation.''}
\end{quote}

\begin{quote} 
\textsc {Article 28: Processor.} \textsl{``the controller shall use only processors providing sufficient guarantees that [...] will meet the requirements of this Regulation.''}
\end{quote}

Together, these articles set the guidelines for security and privacy in a GDPR world. First, article 25 specifies that all systems must be designed, configured, and administered with data protection as a primary goal. Then, article 24 establishes that the ultimate responsibility for the security of all personal data lies with the controller. Lastly, article 28 precludes the controllers from using any processors (in our context, cloud providers) who do not meet the requirements of GDPR.

\vheading{Security in the real-world}. Cloud providers, who act as processors, have been swift in showcasing~\cite{aws-gdpr-ready, gce-gdpr-ready, azure-gdpr-ready} the compliance of their service offerings. However, given the monetary and technical challenges in redesigning the existing systems, many organizations are turning to \emph{reactive security}. This is evident in Amazon's latest security offering, Macie~\cite{amazon-macie}, which employs machine learning techniques to automatically discover, monitor, and protect personally identifiable information on behalf of legacy cloud applications. 


\section{Concluding Remarks}
\label{sec-discussion}

Achieving compliance with GDPR, while necessary, is not trivial. In this paper, we examine how GDPR regulations conflict with the design, architecture, and operation of modern computing systems. Specifically, we illustrate this tussle via \emph{the seven GDPR sins}. The goal of our work is to highlight the challenges of compliance, especially for existing systems. We hope this serves as a starting point for designing privacy-aware systems. 

\vheading{Controversial points}.
Calling any point in the design spectrum a \emph{sin} is bound to be controversial. We acknowledge that no one could have designed systems for regulations that did not exist at the time, and that companies are unlikely to make similar choices in the new environment. However, GDPR is already here, the people are now aware of their privacy rights, and the regulators are vigilant. Thus, we believe that the tone of this paper reflects the gravity of the situation, and the urgency with which the system designers should respond. 

\vheading{Open issues}.
While our exposition focuses on seven systematic violations of privacy and security, there are many other unsavory practices that we have not covered. For example, the design and operation of online behavioral tracking~\cite{behavioral-ads}. Nor have we prescribed any policies or mechanisms towards achieving compliance. Also, the seven practices highlighted here exist due to technical and economical reasons that may not entirely be in the control of individual companies. Thus, solving such deep rooted issues would likely result in significant performance overheads, slower product rollouts, and reorganization of data markets. The equilibrium points of these tussles are not yet clear. 

\vheading{Future directions}.
Given the scope and scale of GDPR, compliance is likely going to be a slow and messy endeavor. So, how should the systems community tackle this problem? Addressing compliance at the level of individual infrastructure components (i.e., compute, storage, and networking) versus at the level of individual regulations will result in different tradeoffs. While the former makes the effort more contained (and suits the cloud model better), the latter provides opportunites for cross-layer optimizations. Another challenging topic is that of testing for compliance: should it be proactive or reactive? How much of the detection and reporting be automated versus manual? How should compliance be priced in the cloud?

We expect the paper to generate interesting discussions at HotCloud. GDPR is not only a comprehensive privacy legislation but the first one at that. As several other nations are in the process of drafting privacy regulations, participation from the systems community would be valuable.

{
\small
\bibliographystyle{plain}
\bibliography{paper}

\begin{thebibliography}{10}

\bibitem{ferpa}
{Family Educational Rights and Privacy Act}.
\newblock {\em Title 20 of the United States Code, Section 1232g}, Aug 21 1974.

\bibitem{hipaa}
{The Health Insurance Portability and Accountability Act}.
\newblock {\em 104th United States Congress Public Law 191}, Aug 21 1996.

\bibitem{google-deletion}
{Data Deletion on Google Cloud Platform}.
\newblock \url{https://cloud.google.com/security/deletion/}, Sep 2018.

\bibitem{amazon-macie}
Amazon {M}acie.
\newblock \url{https://aws.amazon.com/macie/}, Accessed Jan 31 2019.

\bibitem{google-takeout}
Google {T}akeout.
\newblock \url{https://takeout.google.com/}, Accessed Jan 31 2019.

\bibitem{xAI}
David Aha, Trevor Darrell, Michael Pazzani, Darryn Reid, Claude Sammut, and
  Peter Stone, editors.
\newblock {\em Workshop on Explainable Artificial Intelligence}. International
  Joint Conference on Artificial Intelligence (IJCAI), August 2017.

\bibitem{data-is-bacon}
Forsyth Alexander.
\newblock Data is the new bacon. {I}n \emph{IBM Business analytics blog}.
\newblock
  \url{https://www.ibm.com/blogs/business-analytics/data-is-the-new-bacon/},
  Oct 18 2016.

\bibitem{zuckerberg-quote}
Henry Blodget.
\newblock Mark zuckerberg on innovation. {I}n \emph{Business Insider}, Oct 1
  2009.

\bibitem{gdpr-in-numbers}
The European Data~Protection Board.
\newblock {GDPR} in {N}umbers.
\newblock
  \url{https://ec.europa.eu/commission/sites/beta-political/files/190125_gdpr_infographics_v4.pdf},
  Jan 25 2019.

\bibitem{roman-law}
William Buckland and Peter Stein.
\newblock {\em {A text-book of Roman law: From Augustus to Justinian}}.
\newblock Cambridge University Press, 2007.

\bibitem{google-purpose-bundling}
CNIL.
\newblock The {CNIL}’s restricted committee imposes a financial penalty of 50
  million euros against {Google LLC}.
\newblock
  \url{https://www.cnil.fr/en/cnils-restricted-committee-imposes-financial-penalty-50-million-euros-against-google-llc},
  January 21st 2019.

\bibitem{gdpr-data-download}
Kate Conger.
\newblock {How to Download Your Data With All the Fancy New {GDPR} Tools}. {I}n
  \emph{Gizmodo}.
\newblock
  \url{https://gizmodo.com/how-to-download-your-data-with-all-the-fancy-new-gdpr-t-1826334079},
  May 25 2018.

\bibitem{gdpr-programmatic-ad-buy}
Jessica Davies.
\newblock {GDPR mayhem: Programmatic ad buying plummets in Europe}. {I}n
  \emph{Digiday}.
\newblock
  \url{https://digiday.com/media/gdpr-mayhem-programmatic-ad-buying-plummets-europe/},
  May 25 2018.

\bibitem{nytimes-no-ads}
Jessica Davies.
\newblock {After GDPR, The New York Times cut off ad exchanges in Europe}. {I}n
  \emph{Digiday}.
\newblock
  \url{https://digiday.com/media/new-york-times-gdpr-cut-off-ad-exchanges-europe-ad-revenue/},
  Jan 16 2019.

\bibitem{aadhar-refute-breach}
Vidhi Doshi.
\newblock {A security breach in India has left a billion people at risk of
  identity theft}. {I}n \emph{The Washington Post}.
\newblock
  \url{https://www.washingtonpost.com/news/worldviews/wp/2018/01/04/a-security-breach-in-india-has-left-a-billion-people-at-risk-of-identity-theft},
  Jan 4 2018.

\bibitem{gartner-prediction}
Amy~Ann Forni and Rob van~der Meulen.
\newblock Organizations are unprepared for the 2018 {E}uropean {D}ata
  {P}rotection {R}egulation. {I}n \emph{Gartner}, May 2017.

\bibitem{gdpr-explanation}
Bryce Goodman and Seth Flaxman.
\newblock European {U}nion {R}egulations on {A}lgorithmic {D}ecision-{M}aking
  and a {R}ight to {E}xplanation.
\newblock {\em AAAI AI Magazine}, 38(3), 2017.

\bibitem{panera-delay-breach}
Michael Grothaus.
\newblock {Panera Bread leaked millions of customers’ data}. {I}n \emph{Fast
  Company}.
\newblock
  \url{https://www.fastcompany.com/40553518/report-panera-bread-leaked-millions-of-customers-data},
  Apr 3 2018.

\bibitem{uber-breach-payoff}
Mike Isaac, Katie Benner, and Sheera Frenkel.
\newblock {Uber Hid 2016 Breach, Paying Hackers to Delete Stolen Data.} {I}n
  \emph{The New York Times}.
\newblock \url{https://www.nytimes.com/2017/11/21/technology/uber-hack.html},
  Nov 21 2017.

\bibitem{WHI}
Been Kim, Dmitry Malioutov, and Kush Varshney, editors.
\newblock {\em Workshop on Human Interpretability in Machine Learning}.
  International Conference on Machine Learning (ICML), June 2016.

\bibitem{behavioral-ads}
Natasha Lomas.
\newblock {Even the IAB warned adtech risks EU privacy rules}. {I}n
  \emph{TechCrunch}.
\newblock
  \url{https://techcrunch.com/2019/02/21/even-the-iab-warned-adtech-risks-eu-privacy-rules/},
  Feb 21 2019.

\bibitem{noyb-streaming-service-complaints}
Natasha Lomas.
\newblock {Privacy campaigner Schrems slaps Amazon, Apple, Netflix, others with
  GDPR data access complaints.} {I}n \emph{TechCrunch}, Jan 18 2019.

\bibitem{strava-military-leak}
James Quarles.
\newblock {An Update on the Global Heatmap}.
\newblock
  \url{https://blog.strava.com/press/a-letter-to-the-strava-community/}, Jan 29
  2018.

\bibitem{azure-gdpr-ready}
Alym Rayani.
\newblock {Safeguard individual privacy rights under GDPR with the Microsoft
  intelligent cloud.} {I}n \emph{Microsoft 365 Blog}.
\newblock
  \url{https://www.microsoft.com/en-us/microsoft-365/blog/2018/05/25/safeguard-individual-privacy-rights-under-gdpr-with-the-microsoft-intelligent-cloud/},
  May 25 2018.

\bibitem{gdpr-regulation}
General Data~Protection Regulation.
\newblock Regulation ({EU}) 2016/679 of the {E}uropean {P}arliament and of the
  {C}ouncil of 27 {A}pril 2016 on the protection of natural persons with regard
  to the processing of personal data and on the free movement of such data, and
  repealing {D}irective 95/46.
\newblock {\em Official Journal of the European Union}, 59(1-88), 2016.

\bibitem{strava-heatmap}
Drew Robb.
\newblock {Building the Global Heatmap}.
\newblock
  \url{https://medium.com/strava-engineering/the-global-heatmap-now-6x-hotter-23fc01d301de},
  Nov 1 2017.

\bibitem{facebook-view-as-leak}
Guy Rosen.
\newblock {Security Update}.
\newblock \url{https://newsroom.fb.com/news/2018/09/security-update/}, Sep 28
  2018.

\bibitem{gdpr-storage}
Aashaka Shah, Vinay Banakar, Supreeth Shastri, Melissa Wasserman, and Vijay
  Chidambaram.
\newblock {Analyzing the Impact of GDPR on Storage Systems}.
\newblock In {\em USENIX HotStorage}, 2019.

\bibitem{cambridge-analytica}
Olivia Solon.
\newblock {Facebook says Cambridge Analytica may have gained 37{M} more users'
  data}. {I}n \emph{The Guardian}.
\newblock
  \url{https://www.theguardian.com/technology/2018/apr/04/facebook-cambridge-analytica-user-data-latest-more-than-thought},
  Apr 4 2018.

\bibitem{usa-today-no-ads}
Erica Sweeney.
\newblock {Many publishers' EU sites are faster and ad-free under GDPR}. {I}n
  \emph{Marketing Dive}.
\newblock
  \url{https://www.marketingdive.com/news/study-many-publishers-eu-sites-are-faster-and-ad-free-under-gdpr/524844/},
  Jun 4 2018.

\bibitem{pre-gdpr-numbers}
Ed~Targett.
\newblock {6 Months, 945 Data Breaches, 4.5 Billion Records.} {I}n
  \emph{Computer Business Review}.
\newblock \url{https://www.cbronline.com/news/global-data-breaches-2018}, Oct 9
  2018.

\bibitem{move-fast-break-things}
Moshe Vardi.
\newblock {Move Fast and Break Things}.
\newblock {\em Communications of the ACM}, 61(9), 2018.

\bibitem{gdpr-no-right-to-explanation}
Sandra Wachter, Brent Mittelstadt, and Luciano Floridi.
\newblock Why a right to explanation of automated decision-making does not
  exist in the general data protection regulation.
\newblock {\em International Data Privacy Law}, 7(2):76--99, 2017.

\bibitem{gce-gdpr-ready}
Google~Cloud Whitepaper.
\newblock {Google Cloud and the GDPR}.
\newblock Technical report, Google Inc., May 2018.

\bibitem{aws-gdpr-ready}
Chad Woolf.
\newblock {All AWS Services GDPR ready.} {I}n \emph{AWS Security Blog}.
\newblock
  \url{https://aws.amazon.com/blogs/security/all-aws-services-gdpr-ready/}, Mar
  26 2018.

\end{thebibliography}
}

\end{document}